\documentstyle{EuroPhys}

\newif\ifboo \boofalse

\input epsf

\newcommand{\be}{\begin{equation}}
\newcommand{\ee}{\end{equation}}
\def\spose#1{\hbox to 0pt{#1\hss}}
\def\ltapprox{\mathrel{\spose{\lower
3pt\hbox{$\mathchar"218$}}
 \raise 2.0pt\hbox{$\mathchar"13C$}}}
\def\gtapprox{\mathrel{\spose{\lower
3pt\hbox{$\mathchar"218$}}
 \raise 2.0pt\hbox{$\mathchar"13E$}}}
\def\inapprox{\mathrel{\spose{\lower
3pt\hbox{$\mathchar"218$}}
 \raise 2.0pt\hbox{$\mathchar"232$}}}

\begin{document}
\shorttitle{Gabrielli et al., Gravitational 
force distribution in fractal structures}

\title{Gravitational force distribution in fractal structures }
\author{A. Gabrielli \inst{1,2},F. Sylos Labini\inst{1,3},
   and S. Pellegrini\inst{1}}
\institute{
     \inst{1} INFM Sezione di Roma1,  Universit\`a di 
           ``La Sapienza'', Department of   Physics -
           P.le A. Moro 2, I-00185 Roma, Italy\\
      \inst{2} Dipartimento di Fisica, 
           Universit\`a degli Studi ``Tor Vergata'', 
            V.le della Ricerca Scientifica, 1, 00133 Roma, Italy\\
      \inst{3} D\'ept.~de Physique Th\'eorique,
            Universit\'e de Gen\`eve 24, Quai E. Ansermet, 
            CH-1211 Gen\`eve, Switzerland \\
	}
\rec{ }{ }
\pacs{
\Pacs{05}{20$-$y}{Statistical Mechanics}
\Pacs{98}{65$-$r}{Large scale structure of the Universe}
      }
\maketitle

\begin{abstract}
We study the (newtonian) gravitational force distribution
arising from a fractal set of sources.
We show that, in the case of 
real structures in finite samples,  an important role
is played by morphological properties
and finite size effects.
For dimensions smaller than $d-1$ 
(being $d$ the space dimension)
the convergence of the net gravitational
force is assured by the fast decaying of the density,
while for fractal dimension $D>d-1$ the morphological
properties of the structure 
determine the eventual convergence of the force
as a function of distance. We clarify the role
played by the cut-offs of the distribution.
Some cosmological implications are discussed.
\end{abstract}

 The aim of the present paper is to discuss the general properties of the
 gravitational field generated by a finite fractal distribution of field
 sources. This problem is nowadays particularly relevant.
 In fact, there is a general agreement that galaxy distribution exhibits
 fractal behavior up to a certain scale \cite{slmp98,pee93}. 
 The eventual presence of a transition 
 scale towards homogeneity and the exact value of the fractal 
 dimension  are 
 still   matters of debate
 \cite{dav97,pmsl97,wu98}. 
 Moreover it has been observed that
 cold gas clouds of the interstellar medium
 have a fractal structure, with $ 1.5 \le D \le 2$ in a 
 large range of length scales \cite{lar}. 
 For this case 
 the general belief is that the origin 
 of their fractality lies in turbulence.
 However recently \cite{sanchez96} it has been 
 pointed out that self-gravity itself 
 may be the dominant factor in
 making clouds fractal.


 Chandrasekhar \cite{chandra1},   has considered 
 the behavior  of the newtonian gravitational force  
 probability density arising 
 from a  poissonian 
 distribution of sources (stars).  
 In this case, in 
 evaluating the probability distribution  of the force   acting 
 on a particle, one supposes that fluctuations are
 subject to the restriction 
 that a constant average density occurs, i.e. the source
 distribution is spatially 
 homogeneous and the density fluctuations obey to 
 the Poisson statistics.
 Applying the Markov's 
 method, it is possible to compute  
 explicitly the  force probability density,
 known as the Holtsmark's distribution.
  In this case, the probability density of the force $F$
 is given by
 \begin{equation}
 \label{h1}
 W(F) = \frac{H(\beta)}{F_o}
 \end{equation}
 where
 $F_o = (4/15)^{2/3} (2 \pi GM) n^{2/3}$ is 
 the normalizing force 
 ($n$ is the average density of sources,  $M$
 is the mass of each punctual 
 source and $G$ is the gravitational constant), 
 $ \beta = F/F_o$ is an adimensional force and 
 \begin{equation}
 \label{h2}
 H(\beta) = \frac{2}{ \pi \beta}  
 \int_0^{\infty} dx\; \exp[ -(x/\beta)^{\frac{3}{2}}]\;x\;\sin(x) \; .
 \end{equation}
 The main result is that, in the 
 thermodynamic limit ($V \rightarrow \infty$ with $n$ constant),
 the force distribution has a finite first moment and 
 an infinite variance. This divergence is due to the possibility of
 having two field sources arbitrarily nearby. 
 The  approximate solution given by the 
 nearest neighbor (n.n.) approximation (i.e. by considering only the 
 effect of the n.n. particle) 
 and the exact Holtsmark's distribution (Eq.\ref{h2}) agrees 
 over
 most of range of $F$ (see Fig.1).
 The region where they differ mostly is when $F \rightarrow 0$. 
 This is due to 
 the fact
 that a weak field arises
 from a more or less 
 symmetrical distribution 
 of points, and hence the n.n. approximation fails. 
 Therefore 
 in the strong field  limit we may
 neglect the contribution to the force of far away 
 points,
 because the main contribution is due to $r \rightarrow 0$ 
 (i.e. it comes from 
 the n.n). The root mean square of the force 
 is   divergent as in the case of the n.n.  approximation, and 
 this is due to the fact that the limit $r \rightarrow 0$ is allowed.
 If there exists a lower cut-off (see below) this divergence is
 erased out. However, due to isotropy, no divergence problem arises 
 from the faraway sources even in the limit $r\rightarrow \infty$.

 
 The derivation of Chandrasekhar cannot be easily extended 
 to the case of fractal distributions  because, in such   structures,
 fluctuations 
 are characterized by   long-range correlations.
   A fractal is an instrisic critical structure
 characterized by self-similarity. The scale-invariant
 properties of such distributions are described
 by slow power law decreasing density-density correlation functions whose 
 exponents are
 the main characteristic quantities to be studied
 (see \cite{man77} and \cite{slmp98}).    
 In this situation an analytic treatment 
 of the force distribution becomes very difficult.  
 Vlad \cite{vlad}  developed a functional integral approach 
 for evaluating the stochastic properties of vectorial 
 additive random fields 
 generated by a variable number of point sources obeying to
 inhomogeneous Poisson statistics.
 Then he applied these results to the case of the gravitational
 force  generated by a 
 fractal distribution of
 field sources under some strong approximations.
 In particular, in order to compute analytically the 
 force probability density, Vlad \cite{vlad} has not considered
 the presence of intrinsical fluctuations in the 
 space density. In fact, in the computation of the 
 density from a single point one expects to see 
 deviation from the average scaling behavior, which
 are present at any scale (see below).
 Such a situation occurs in any real fractal structure
 and can be quantified by studying the $n$-point correlation
 functions \cite{ball}.
 Instead, the derivation 
 has been done under the assumption that 
 the $n+1$-point correlation function 
 (where the $(n+1)^{th}$ point is the occupied origin) 
 can be written as
$ g(\vec{x_1},\vec{x_2},...,\vec{x_n}) 
 \propto g(\vec{x_1}) g(\vec{x_2})...g(\vec{x_n}) \;,
$
 where $g(\vec{x})$ is the two-point correlation function.
 Such an approximation is not
 adequate to describe the effect of morphology
 and hence in real cases the situation is  quite different.
 Instead of Eq.\ref{h2}, Vlad found that 
 the probability density of the absolute value $F$ (generalized 
 Holtsmark's distribution) 
 of the field intensity is equal to 
 \begin{equation}
 \label{h3}
 H(\beta,D) = \frac{2}{\pi\beta} \int_0^{\infty} 
 dx \; \exp[-(x/\beta)^{\frac{D}{2}}]\;x\;\sin(x)\; .
 \end{equation}
 In this case $F_o = (4/15)^{2/D} (2 \pi GM) (DB/(4\pi))^{2/D}$,
 where    $B$ is a constant characterizing 
 the average mass in the unitary sphere and $ \beta = F/ F_o$. 
 The main change due to the  fractal structure is that 
 the scaling exponent in Eq.\ref{h3} 
 is $D/2$ rather than $3/2$. Hence in this case the tail of
 the probability density has a 
 slower decay than in the homogeneous case (see Fig.1).
 This means that the variance of the force is larger for $D <3$ 
 than for the $D=3$.
 As we discuss below, the case $D<2$ is rather well 
 described by Eq.\ref{h3}: this is not the case
 for $D>2$ where the $n-th$ points correlations
 must be taken into account
 in the case of real structures.  
 An important limit is the strong field one  ($F \rightarrow \infty$). 
 In this case it is possible to show that the 
 force distribution of Eq.\ref{h3}
 can be  reduced to the one derived under the n.n. 
 approximations:
\begin{equation}
 \label{f4}
 W_{nn}(F) d F = \frac{D B}{2} (G M)^{\frac{D}{2}} 
 F^{-\frac{D+2}{2}}\\ 
 \exp \left( - B (GM)^{\frac{D}{2}}
 F^{-\frac{D}{2}} \right) d F \;.
\end{equation}
 The n.n. approximation is good for
 $F \gg F_1$, where 
 $F_{1}= \left(B\right)^{ \frac{2} {D} }
 \cdot GM $. From Eq.\ref{f4},
 in the fractal case, as in the homogeneous
 one, the   divergences of the   force moments (the second for $D\le 3$ and 
 the first for $D\le 2$)
 are due only to the fact that two field sources can be arbitrarily close.
 If we impose that a lower cut-off $\Lambda >0$ (minimal distance between
 the sources) exists, these divergences 
 are erased out. 
 According to the approximations previously mentioned,
 far away points do not produce any divergence neither in the 
 average force nor in its variance. This is related 
 to the fact that if the volume involved is large 
 enough with respect to $\Lambda$
 the distribution becomes spherically symmetric  in a statistical sense
 and the anisotropies become negligible.
 
In the n.n. approximation 
 (Eq.\ref{f4}), we obtain for the average force
$ \left\langle F \right\rangle \sim (1/\Lambda)^{2-D}$.
 Hence  in the limit
 $\Lambda \rightarrow 0$, $\left\langle F \right\rangle$ is finite for 
 $D>2$, while
 it is infinite for $D<2$. The force r.m.s value can be computed 
 in the same way and we readily obtain
$ \left\langle F^2 \right\rangle \sim (1/\Lambda)^{4-D}$
 so that,
 in the limit
 $\Lambda \rightarrow 0$,  $\left\langle F^2 \right\rangle$
 is divergent for any possible value of the fractal dimension such 
 that $D \le 3$. 
 Therefore  
 the r.m.s. value of the force strongly depends
 on the  lower cut-off $\Lambda$.
 Consequently, in this context, 
 we expect to see large fluctuations from the average
 value changing the origin point on different masses in the 
 sample.
 This kind of
 divergence is due to the fact that there is no restriction on the 
 distance $\Delta r$ between two neighbor field sources, i.e. we may
 have
 $\Delta r \rightarrow 0$. To assume $\Lambda>0$ is quite reasonable for 
 almost every
 physical problem and in particular for the case of
 galaxy distribution. In fact, if two galaxies were too close, they
 should be  subjected to tidal interactions and  then  should   
 form a binary system. Such a situation goes beyond the 
 scope of this paper.

 As we have already mentioned,  the Vlad's result has been obtained
 under an assumption which is equivalent 
 to the condition that the mass-length relation (m.l.r.) $N(<r)$
 from a single point is {\em exactly} a power law function:
 \begin{equation}
 \label{i2}
 N(<r;\hat{\phi}, \Omega) =
 B \frac{\Omega}{4 \pi} r^D \; .
 \end{equation}
 Eq.\ref{i2} gives 
 the mass contained in a portion of a sphere of radius $r$, 
 being a generic fixed point at the origin, 
 which covers a solid angle $\Omega$ in the 
 direction $\hat{\phi}$.
 From this equation we can derive directly 
 the number density in the same 
 volume by dividing it for the volume itself
 (in $d=3$)
 \begin{equation}
 \label{n}
 n(<r;\hat{\phi}, \Omega) =
 \frac{B}{4\pi} r^{D-3}
 \end{equation}
 It is important to notice, instead, that 
 the m.l.r. {\em has} a genuine power law behavior
 only when it is averaged over all the points $\vec{x}$
 of the structure,  obtaining the so-called two-point correlation 
function (tpcf).
 From a single point there can be important 
 intrinsical fluctuations
 around  the power law behavior \cite{ball}. 
 In fact, since a fractal is an intrinsically 
 critical system with strong correlations
 on all  scales, we expect that the 
 density from one point shows
 large fluctuations,   with respect to the tpcf, changing  
$r$ with fixed  $\hat{\phi}$ or vice-versa.
 In general these intrinsic fluctuations can be 
 characterized as log-periodic corrections to scaling \cite{ball}.
 These fluctuations can be present at any scale 
 (i.e. whatever is $r$) and
 they can be angularly correlated, that is  the quantity 
 \begin{equation}
 \label{dn}
 dn(|\hat{\phi}_1-\hat{\phi}_2|;r)=
 \left\langle(n(<r;\hat{\phi}_1,\delta \Omega)-
 n(<r;\hat{\phi}_2,\delta \Omega))^2 \right\rangle
 \end{equation} 
 depends on the angular distance between $\hat{\phi}_1$ and 
 $\hat{\phi}_2$. 
 As we show below,
 these fluctuations can play a crucial role 
 in the determination of the force distribution.
 In order to consider, in a simple way, the
 important effect of morphology, 
 we study  the r.m.s force for a generic random fractal 
 distribution in the three dimensional Euclidean Space.
 We consider explicitly the existence of 
 a minimal distance (i.e. lower cut-off) $\Lambda$ and a maximal 
 distance (i.e. upper cut-off) $R_s$ from the origin, and 
 then we  derive the 
 asymptotic behavior for $\Lambda \rightarrow 0$
 and $R_s \rightarrow \infty$.
 Let $\vec{F}$ be the total force acting on a particle
 due to all the other   field sources having 
 the same unitary mass
 ($GM=1$). We consider the contribution
 to the force by all the field sources in the 
 spherical shell $[\Lambda,R_s]$
 \begin{equation}
 \label{df1}
 \vec{F}(\Lambda, R_s ) = \sum_{\Lambda \le |\vec{r}_i| \le R_s}^{i=1,N}
 \frac{\vec{r}_i}{r_i^3} 
 \end{equation} 
 where $N$ is the total number of the sources in the shell considered 
 and $\vec{r}_i$ is the position
 of the $i-th$ source with respect to the one in the origin.
 Under the assumption that the mass distribution is statistically isotropic,
 we have
 \begin{equation}
 \label{df2}
 \lim_{R_s/\Lambda \rightarrow \infty}\left| 
 \left\langle \vec{F}(\Lambda, R_s ) 
 \right\rangle\right| = 0 \; .
 \end{equation} 
 The statistical average of the force square modulus
 is given by:
 \begin{equation}
 \label{df4}
 \left\langle |\vec{F}(\Lambda, R_s )|^2 \right\rangle = 
 \left\langle  F(\Lambda, R_s )^2 \right\rangle = 
 \left\langle
 \sum_{\Lambda \le |\vec{r}_i| \le R_s}^{i=1,N} \frac{1}{r_i^4}\right\rangle  
 + \left\langle \sum_{\Lambda \le |\vec{r}_i| \le R_s}^{i \ne j=1, N} 
 \frac{\cos(\theta_{ij})}{r_i^2r_j^2} 
 \right\rangle
 \end{equation} 
 where $\theta_{ij}$ is the angle formed by $\vec{r}_i$ and $\vec{r}_j$.
 Here the average $<...>$ is made over an ensemble
 of different realizations of the sample or,
 when this is not possible,  over all the
 possible origins in a single sample.
 Obviously, 
 the first term is simple to  analyze, while the second one deserves
 a more careful discussion because it involves the 
 three-point correlation function. 
 The first term  can be written as  
 \begin{equation}
 \label{df5}
 \left\langle \sum_{\Lambda \le |\vec{r}_i| \le R_s}^{i=1,N} \frac{1}{r_i^4}
 \right\rangle = D B \int_{\Lambda}^{R_s}
 dr \frac{r^{D-1}}{r^4} 
 =\frac{DB}{4-D} \frac{1}{\Lambda^{4-D}} 
 \left( 1 - \left( 
 \frac{\Lambda}{R_s} \right)^{4-D} 
 \right) \; .
 \end{equation}
 In Eq.\ref{df5} we have considered that $<...>$ leaves only the power law 
 behavior in the m.l.r. In fact we can write 
 $\sum_{i=1}^{N} \frac{1}{r_i^4}=\int_{\Lambda,R_s}d^3 r n(\vec{r})/r^4$, 
 where $n(\vec{r})=\sum_{i=1}^{N}\delta(\vec{r}-\vec{r_i})$
 is the number density, which can 
 be different from a power law and show persistent 
 fluctuations around 
 its average behavior, and $<...>$ erases 
 these fluctuations as well as  the dependence
 from the direction of $\vec{r}$.
 Analogously, 
 the second term in Eq.\ref{df4}  can be written as
 \begin{equation}
 \label{df6}
 \left\langle 
 \sum_{\Lambda \le |\vec{r}_i| \le R_s}^{i \ne j=1, N} 
 \frac{\cos(\theta_{ij})}{r_i^2r_j^2}  
 \right\rangle=
 \int_{V} \int_{V} d^3 r_1  d^3 r_2 p(\vec{r}_1,\vec{r}_2)  
 \frac{\cos(\theta_{12})}{r_1^2 r_2^2} 
 \end{equation}
 where $p(\vec{r}_1,\vec{r}_2)d^3 r_1  d^3 r_2  $ 
 is equal to $N(N-1)$ times
 the probability to have a 
 particle in the infinitesimal volume $d^3 r_1 $
 and, at the same time, another in the 
 volume $d^3 r_2$, i.e. the number of pairs
 in which the first particle is in $d^3 r_1 $
 and the second in  $d^3 r_2$.
 The function $p(\vec{r}_1,\vec{r}_2)$
 is strictly related to the
 three-point correlation function in which the first point is
 the origin and the others are $\vec{r}_1$ and $\vec{r}_2$. 
 Such a function is usually very difficult to be computed
 for a generic fractal structure 
 \cite{ball}. 
 Note that  $p(\vec{r}_1,\vec{r}_2)
 \sim \left\langle n(\vec{r}_1,\vec{r}_2)\right\rangle$ where 
$ n(\vec{r}_1,\vec{r}_2) = \sum_{i,j} \delta(\vec{r_i}-\vec{r_1})
\delta(\vec{r_j}-\vec{r_2})\;.
$
 We expect that $p(\vec{r}_1,\vec{r}_2)$ 
 depends only on $r_1$ and $r_2$ and the 
 angle $\theta$ between the two directions because of the average 
 $<...>$.
 We make the following approximation,  
 by using the tpcf and 
 the angular correlation function   
 \begin{equation} 
 \label{df7}
 \int_{\theta \le |\vec{\Omega}_1 - \vec{\Omega}_2| \le \theta + d\theta} 
 p(\vec{r}_1,\vec{r}_2) d \Omega_1  d \Omega_2 
 \simeq (4 \pi)^2 \left\langle n(r_1)\right\rangle \left\langle n(r_2)
 \right\rangle  \Gamma\left(\theta,\frac{r_1}{r_2}\right) d \theta 
 \end{equation}
 where $\Gamma(\theta,r_1/r_2)$ is the probability 
 that the angle between $\vec{r}_1$ and  $\vec{r}_2$
 is in the range $[\theta, \theta + d \theta]$, conditioned to the fact that 
 the distances of the two sources are
 $r_1$ and  $r_2$.  
 The  dependence of  $\Gamma(\theta,r_1/r_2)$ on the 
 ratio $r_1/r_2$ is 
 in general a peculiar property of the  fractal
 studied.
 This problem has been studied in detail  by \cite{ball}. 
 In what follows, we neglect 
 the dependence on the ratio $r_1/r_2$,
 which can be treated as a perturbation \cite{ball}.
 This approximation consists in substituting 
 $\Gamma(\theta,r_1/r_2)$ 
 with the angular correlation function $\Gamma(\theta)$ 
 normalized to unity.
 $\Gamma(\theta)$ is 
 obtained by the analysis of the angular
 projection of the sample.

 Under this approximation we can write Eq.\ref{df6} as
 \begin{equation}
 \label{df10}
 \left\langle 
 \sum_{\Lambda \le |\vec{r}_i| \le R_s }^{i \ne j=1, N}
 \frac{\cos(\theta_{ij})}{r_i^2r_j^2}  
 \right\rangle= \frac{(DB)^2  C(D)}{(2-D)^2} \frac{1}{\Lambda^{4-2D}} 
 \left( 1 + \left(\frac{\Lambda}{R_s}\right)^{4-2D} -
 2 \left(\frac{\Lambda}{R_s}\right)^{2-D} \right)
 \end{equation}
  where
$ C(D) = \int_0^{\pi}   \Gamma(\theta) \cos(\theta) d \theta \; .
$
 The presence of fluctuations in the m.l.r. from one 
 point are crucial in determining the behavior of $\Gamma(\theta)$
 and then the convergence properties of $\left\langle F^2 \right\rangle$.
 To clarify this point, consider the quantity 
 $dn(|\hat{\phi}_1-\hat{\phi}_2|;r)$
 in Eq.\ref{dn}. It is simple to 
 show analytically that this quantity is strictly related 
 to $\Gamma(\theta)$ where $\theta=|\hat{\phi}_1-\hat{\phi}_2|$ 
 is the angular 
 distance between $\hat{\phi}_1$ and $\hat{\phi}_2$.
 At this point is convenient to distinguish the cases $D<2$ and
 $D\ge 2$.
 For $D<2$, even in absence of fluctuations in the m.l.r., we have $C(D)>0$ 
 (the effect of fluctuations is only a change in the
 non-vanishing value of $C(D)$), while for $D\ge 2$, 
 $C(D)>0$ only because of these fluctuations. In the latter case and 
 in absence of 
 fluctuations, one would have a case very similar to the homogeneous one
 \cite{man77}  in
 which $\Gamma(\theta)$ is constant and therefore $C(D)=0$.
 
 Consequently, for $D<2$, the leading behaviors of 
 $\left\langle F^2 \right\rangle$ are:
 \begin{equation}
 \left\{
 \begin{array}{ll}
 \left\langle F^2 \right\rangle\simeq\alpha(B,D)\Lambda^{-(4-D)} & for\;
 \Lambda\rightarrow 0 \\
 & and \; R_s finite\\
 \left\langle F^2 \right\rangle \simeq F^2_{\infty}-2\beta(B,D)
 \left(\frac{\Lambda}{R_s}
 \right)^{2-D}& for \; \frac{R_s}{\Lambda}\rightarrow\infty \\
 & and \; \Lambda>0\\
 \end{array}
 \right. 
 \label{force1} 
 \end{equation}
 In Eq.\ref{force1}, $\alpha(B,D)=BD/(4-D)$ , 
 $\beta(B,D)= (DB)^2C(D)/[(2-D)^2\Lambda^{4-2D}$ and 
 $F^2_{\infty}=\alpha(B,D)\Lambda^{-(4-D)}+\beta(B,D)$ is the finite 
 asymptotic value of $\left\langle F^2 \right\rangle$ for 
 $R_s/\Lambda\rightarrow\infty$.
 Obviously for intermediate ranges of $R_s/\Lambda$, 
 other powers of it in Eq.\ref{df5} and Eq.\ref{df10} are important.
 
 The case $D>2$ is very different and the effect of eventual fluctuations
 in the m.l.r. from one point can be crucial for the convergence of
 $\left\langle F^2 \right\rangle$ in the limit 
 $R_s/\Lambda \rightarrow\infty$.
 As a first step let us suppose that the m.l.r. from one 
 point is a pure power law.
 In this case the function $\Gamma(\theta)$ is expected to be a constant and
 then $C(D)=0$.
 As a consequence we have that the second term in Eq.\ref{df4} is 
 equal to zero and  $\left\langle F^2 \right\rangle$ is simply given by
 \begin{equation}
 \label{force2}
 \left\langle F^2 \right\rangle=
 \frac{DB}{4-D} \frac{1}{\Lambda^{4-D}} 
 \left( 1 - \left( 
 \frac{\Lambda}{R_s} \right)^{4-D} 
 \right) \; .
 \end{equation}
 This quantity diverges for $\Lambda\rightarrow 0$ as
 $\Lambda^{-(4-D)}$ and for $\Lambda>0$ it converges for 
 $R_s/\Lambda\rightarrow\infty$.
   
 In the case in which 
 fluctuations around the power-law relation in the m.l.r. are present 
 (and this is the more general case), the behavior is more complex.
 In this case $\Gamma(\theta)$ is not anymore constant,
 since we expect that the effect of these fluctuations is 
 to introduce some angular correlations at least at small angles.
 The reason is that these fluctuations are due to the sequence of voids and
 clusters of masses in a fractal sample.
 These sequences are not the same in different 
 directions, but they are angularly correlated and the average
 $<...>$ does not erase these correlations.
 In such a case $\left\langle F^2 \right\rangle$ is given by
 \begin{equation}
 \label{force3}
 \left\langle F^2 \right\rangle=
 \frac{DB}{4-D} \frac{1}{\Lambda^{4-D}}
 \left( 1 - \left(
 \frac{\Lambda}{R_s} \right)^{4-D}
 \right) 
 +\frac{(DB)^2  C(D)}{(D-2)^2} \frac{1}{\Lambda^{4-2D}}
 \left( 1 + \left(\frac{R_s}{\Lambda}\right)^{2D-4} -
 2 \left(\frac{R_s}{\Lambda}\right)^{D-2} \right)
 \end{equation}
 From Eq.(\ref{force3}) 
 we can see that, for $R_s$ finite, $\left\langle F^2 \right\rangle$
 diverges as $\Lambda^{-(4-D)}$ (as in the case $D<2$) 
 when $\Lambda\rightarrow 0$, 
 and for $\Lambda>0$ it diverges 
 as $(R_s/\Lambda)^{2D-4}$ in the limit
 $R_s/\Lambda\rightarrow \infty$.
 This last behavior is in striking contrast 
 with the Vlad's one for the same values of $D$;
 this is due to the fact that as mentioned Vlad 
 does not consider the fluctuations in
 the m.l.r. from a single point.
 At intermediate scales the behavior 
 is more complex and depend on the values of the 
 parameter $D$, $B$ and the value of $C(D)$.

 We have tested the results obtained in Eq.20 by numerical 
 simulations. We have generated 
 artificial fractals by means of different algorithms:
 random cantor set, levy flight and random walk \cite{man77}.
   With these algorithms  the fractal dimension can be tuned 
appropriately (i.e. the tpcf), while the morphological properties of 
the structure
(and hence the higher order correlations) are 
determined by the kind of iterative process choosen.
For this reason the simulations do not
have the same morphology of real galaxy distribution.
A more realistic study for the case of galaxy structures
is in progress, and here we are   concerned with 
a more    general problem. 
 The agreement with Eq.20 is rather good
 for both cases $D<2$ and $D>2$. We stress again 
 that the divergence of Eq.20 as a function of 
 $R_s$ is strictly related to 
 the morphological properties of 
 the structure considered.

 The clarification of the properties
 of the gravitational field in fractal 
 structures is particularly relevant for the 
 studies of the matter distribution 
 in the universe. It is commonly believed 
 that the peculiar velocity field
 (derived as a distortion from a pure 
 linear Hubble law) is generated by
 local matter inhomogeneities in the  
 distribution of galaxies \cite{strauss}.
 A standard hypothesis in this framework
 is that, on large enough scale, 
 the linear perturbation theory holds,
   as density fluctuations are small enought,
 and by using such an approximation 
 one may ``reconstruct'' the peculiar
 velocity field from the matter one 
 or vice-versa. Clearly, if matter distribution
 shows fractal properties, 
 the linear theory fails. A related problem
 consists in the computation 
 of the net gravitational force,     due to 
 the  local inhomogeneities of matter distribution, 
 acting on our galaxy (the dipole) \cite{strauss}.
 Here we have shown that
 the eventual convergence or divergence
 of the dipole is related to
 subtle morphological properties
 of the sources distribution   and not
 only to the prtoperties of the tpcf.
 In the universe we observe 
 an irregular distribution of galaxies
 characterized by fractal properties \cite{slmp98},
 and hence its relationship with the peculiar
 velocity field   must be much more complex
 than the one expected in the linear theory \cite{strauss}.
 Such a situation deserves further studies,
 and in this paper we have considered 
 some basic problems related to it. 
 \bigskip

{\bf Acknowledgments} 
 We would like to warmly thank L. Pietronero and M. Montuori
 for useful comments and remarks.
 We  also  thank  Y. Baryshev, H. de Vega, R. Durrer, 
 M. Joyce and N. Sanchez
 for enlighting discussions.
 A.G. and F.S.L. acknowledge the support of the  
 EEC TMR Network  "Fractal structures and  
 self-organization"  
 \mbox{ERBFMRXCT980183}.


\begin{figure}[h]
\epsfxsize=6cm
\epsffile{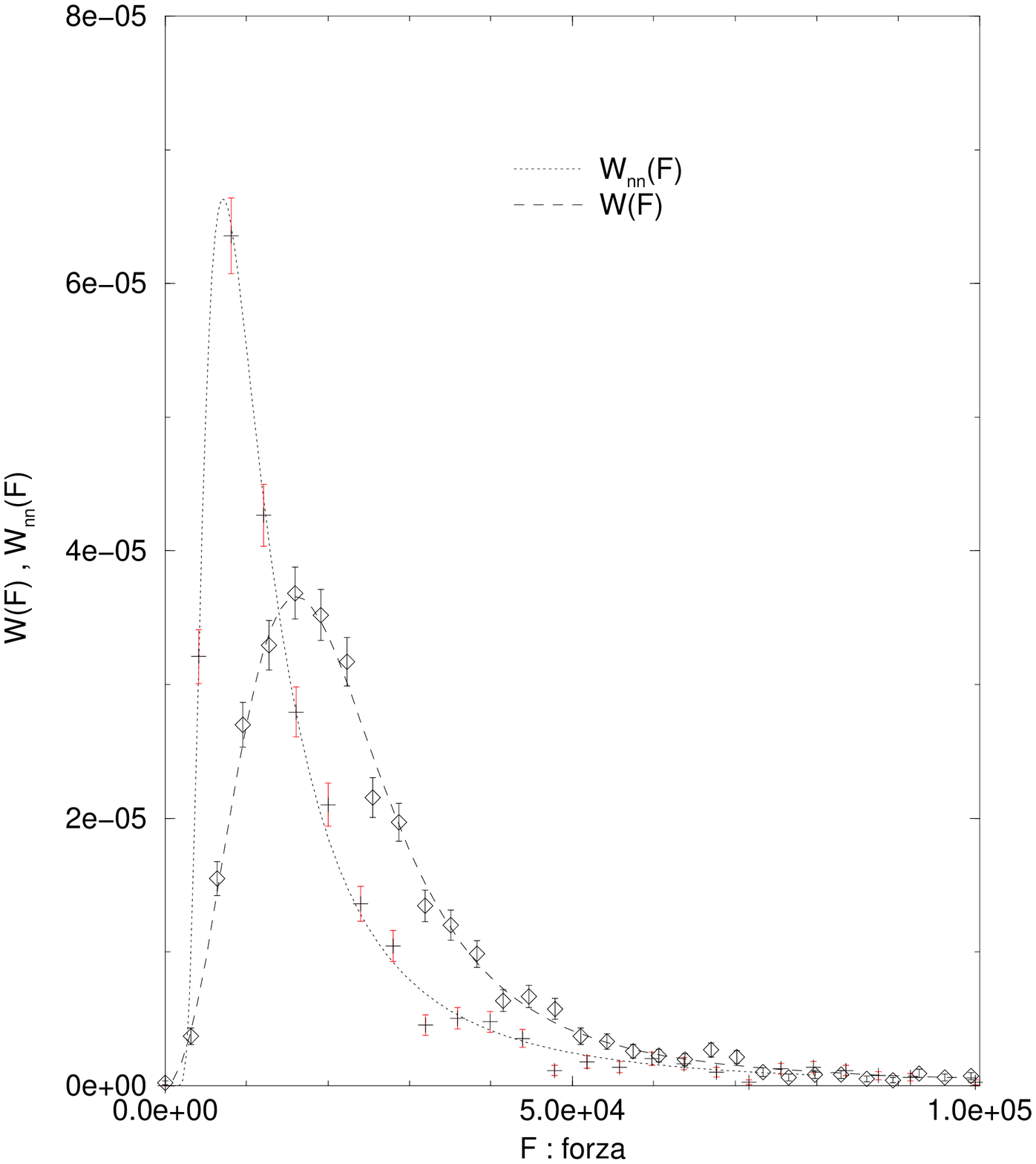}
\epsfxsize=6cm
\epsffile{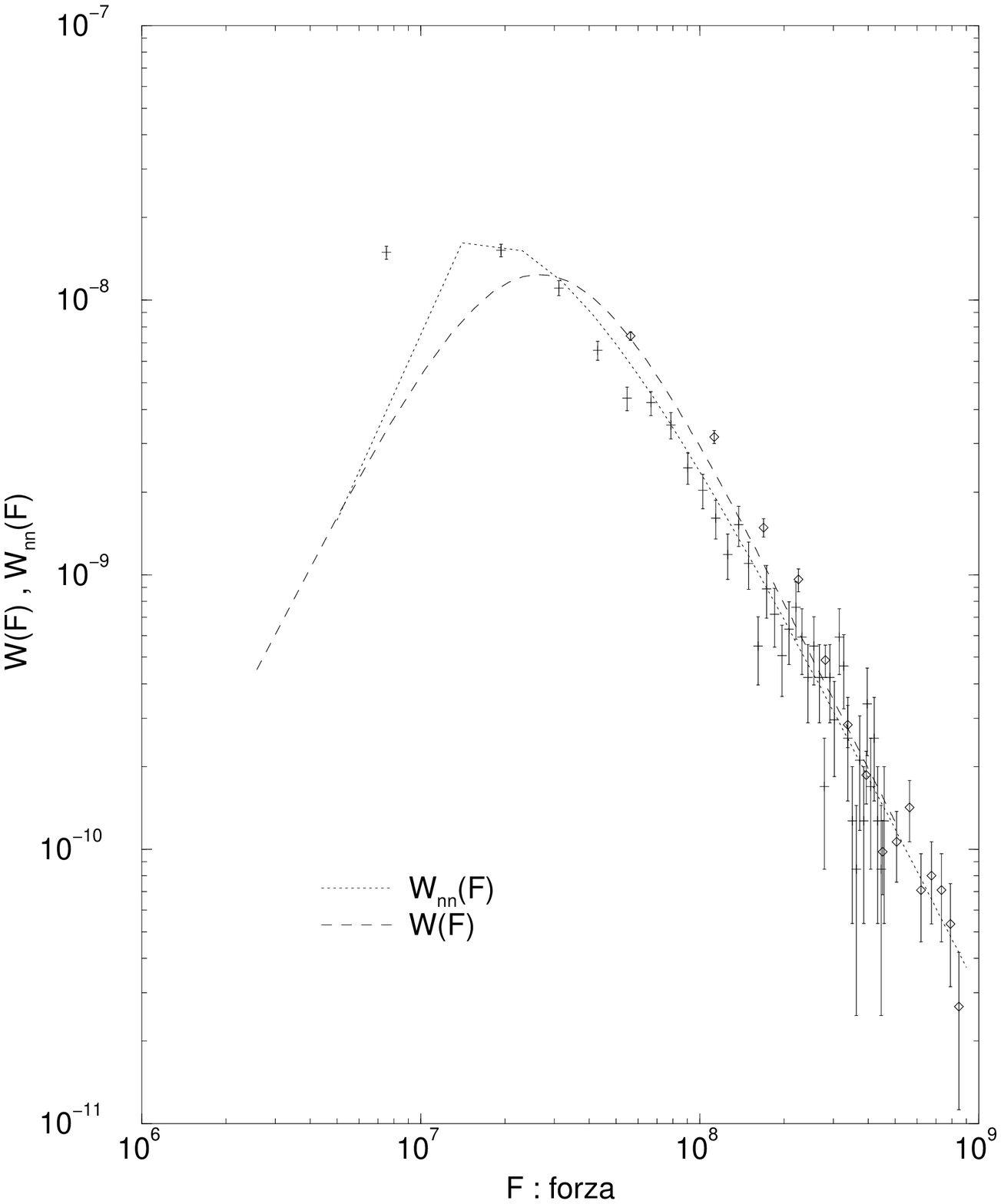}
\caption{(Upper panel) Force distribution due the n.n.
(crosses) and due to all the field sources (diamonds) for 
an homogeneous sample with $n=N/V=2.39 \cdot 10^4$ in $d=3$. 
The dotted line represents
the force distribution $W(F)$ computed by the n.n. approximation, 
while the dashed line is the Holtsmark's distribution (Eq.2). 
The agreement is quite
satisfactory at strong field, and there is a deviation at weak field.
(Bottom Panel) The same for a fractal with $D=1.98$ generated by a 
random walk in $d=3$. In this case we have used the generalized
Holzmark distribution and its n.n. approximation for the 
fits (dotted and dashed lines)}
\end{figure}

\begin{figure}[h]
\epsfxsize=14cm
\epsffile{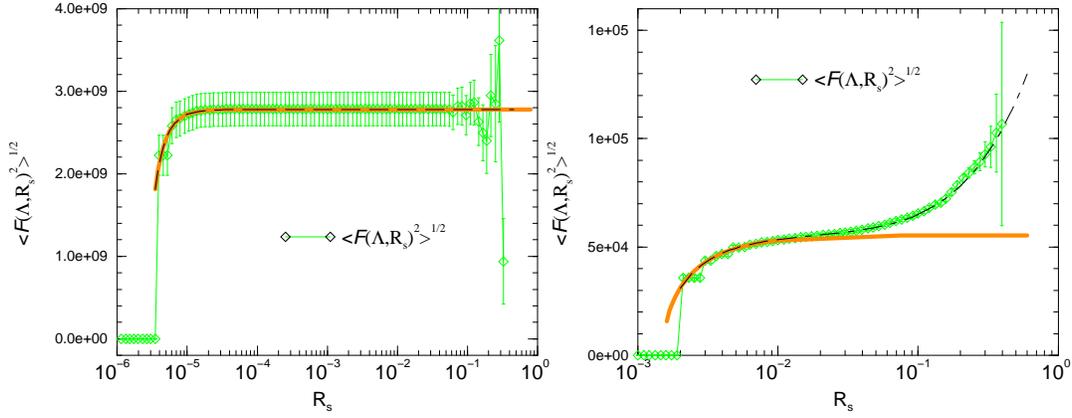}
 \caption{ 
(Left Panel) Behavior of 
$\sqrt{\langle F(\Lambda,R_s)^2\rangle}$ 
as a function of 
$R_s$   for a fractal
with  $D=1.5$ generated by the random cantor set algorithm.
The dashed line represents the first term in Eq.20, while 
the solid line both contributions in Eq.20.
The agreement is rather good. 
(Right Panel)
The same for a fractal with $D=2.7$. 
The effect of the angular correlations  is
now important and determine the divergence of
$\sqrt{\langle F(\Lambda,R_s)^2\rangle}$.
The error bars have been computed over an average 
of 5 different realization of the same structure.
}
 \end{figure}

\end{document}